# A Methodology for Assessing the Environmental Effects Induced by ICT Services

Part II: Multiple Services and Companies

Pernilla Bergmark*
Ericsson Research
Ericsson
Stockholm, Sweden
pernilla.bergmark@ericsson.com

Vlad C. Coroamă*
Department of Computer Science
ETH Zurich
Switzerland
vlad.coroama@inf.ethz.ch

Mattias Höjer
Department of Sustainable Development, Environmental Science and Engineering
KTH Royal Institute of Technology, Sweden
hojer@kth.se

Craig Donovan
Ericsson Research
Ericsson
Stockholm, Sweden
donovan.craig@gmail.com

## ABSTRACT

Information and communication technologies (ICT) can make existing products and activities more efficient or substitute them altogether and could thus become crucial for the mitigation of climate change. In this context, individual ICT companies, industry organizations and international initiatives have started to estimate the environmental effects of ICT services. Often such assessments rely on crude assumptions and methods, yielding inaccurate or even misleading results. The few existing methodological attempts are too general to provide guidance to practitioners. The starting points of this paper are i) a high-level standard from the European Telecommunication Standardisation Institute (ETSI) and the International Telecommunication Union (ITU), and ii) its suggested enhancements for single service assessment outlined in *A Methodology for Assessing the Environmental Effects Induced by ICT Services Part I: Single services (Part I* in short*)*. Building on the assessment of single services, the current article identifies and addresses shortcomings of existing methodologies and industry practices with regard to multiple services assessment. For a collection of services, it addresses the goal and scope definition, the so-far ignored aggregation of effects among several services, and the allocation between several companies contributing to one or more services. The article finally brings these considerations together with those of *Part I* into a workflow for performing such assessments in practice.

* Both authors contributed equally to this paper

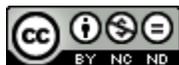

## 1 INTRODUCTION

As outlined in *A Methodology for Assessing the Environmental Effects Induced by ICT Services Part I: Single services (Part I* in short [1]), to limit global warming to 1.5-2 degrees above preindustrial levels, humanity needs to drastically reduce its greenhouse gas (GHG) emissions every decade [2]. Information and communication technologies (ICT) are often envisioned as key enablers of such reductions. They can achieve this by, for example, substituting resource-intensive activities through ICT services – such as replacing conference travel through virtual connections that can entirely [3] or partially [4] virtualise conferences – or by making existing processes more efficient, for example different management services [5].

*Part I* [1] further described how sector-level claims referring to this potential have been put forward by industry associations [6-8], and also by large international bodies such as the European Commission [9, 10], OECD [11], and even the WWF [12-14]. In the wake of these initiatives, individual ICT companies such as British Telecom [15], Telstra [16] and AT&T [17] made efforts to evaluate their own contribution as well. More recently, Mission Innovation, an initiative to promote global clean energy innovation that connects 23 countries plus the EU, presented a framework for avoided emissions which includes ICT solutions within its scope and is intended to help decision makers and investors to support and accelerate innovation of low carbon solutions [18]. In particular, it suggests that investors seeking to change their portfolio profiles must be able to identify products and services which can con-tribute positively to the decarbonization of society, not only those with high footprints. The framework argues that induced effects should thus in the long run be included in company accounting.

Current methods and estimates, however, often rely on crude assumptions and methods, as also acknowledged by some of these initiatives. Moreover, these estimates typically focus exclusively on the potential benefits, ignoring possible negative effects (other than the footprint of ICT itself). A new and accurate methodology thus needs to be developed in order to establish a more credible and consistent fact base. Beyond supporting the scientific discourse, such methodology is needed if such estimates are to be used for business and investment decisions.

To provide for more rigorous assessments, this article proposes methodological guidelines for the assessment of the induced effect of multiple ICT services. It thereby expands *Part I*, which addressed the assessment of single ICT services. Adding to established standards, the article thus undertakes a first step towards a more comprehensive methodology for assessing the environmental effects induced by multiple ICT services and companies beyond their direct footprint. The article presents and categorizes the assessment challenges and reveals common flaws with regard to assessment of multiple services in existing industry claims. Subsequently, it then proposes enhanced assessment principles for multiple services and for allocating their effects to the company-level.

The article is structured as follows: Section 2 summarizes the terminology, which was introduced in more detail in *Part I* [1]. Section 3 introduces the methodological basis and the contributions of the article. Section 4 analyses the assessment challenges from a multi-service perspective and proposes suitable solutions. Section 5 brings *Part I* [1] and *Part II* (this paper) of the methodology together, summarizing our proposed assessments guidelines for both single and multiple services, and introducing a corresponding assessment workflow for company assessments; Section 6 discusses the limitations of our work and suggests directions for further research.

## 2 TERMINOLOGY

*Part I* [1] described the two main categories of environmental impacts associated with ICT:

  A. The impacts associated with the *direct (environmental) footprint*, which include raw materials acquisition, production, use, and end-of-life treatment.
  B. A vast collection of subtler environmental effects induced by the usage of ICT infrastructure and devices, ranging from the short-term impacts of an ICT service to the long-term socio-economic consequences of ICT deployment in general.

As argued in *Part I* [1], the direct footprint is always an environmental burden, while effects in the second category can be environmentally either positive or negative. *Part I* of the paper presents in detail the terminology from the literature, and places our work in its context. As stated there, the scope of our analysis coincides with the 'second-order effects' from [19, 20] (as well as the 'application' category from [21]), taken together with the direct rebound effect from the 'other' category of [19, 20]. In the remainder of this paper – as in *Part I* – we refer to these effects as *induced (envi-ronmental) effects*, keeping in mind that this definition does not cover all the long-term behavioural and structural changes. While the induced effects can be both positive and negative, when referring specifically to positive induced effects, we use synonymously the terms *enablement* or *enabling effect*, in line with the literature. We further refer to two mechanisms behind induced effects, *substitutions* and *optimizations*.

## 3 METHODOLOGICAL BASIS AND CONTRIBUTIONS

Our methodology discussion around the assessment of the induced effect of multiple ICT services starts from a standard jointly developed by ETSI [19] and ITU [20] – from now on referred to as the 'ETSI/ITU standard' – and our methodological developments for single-service assessment, as introduced in *Part I* [1]. Together, they yield the picture depicted in Fig. 1, which combines the standard's general guidelines and our enhancements for single-service assessment: substitution and optimization, time perspective, baseline setting, case studies versus models, extrapolation from case studies, and the influence of direct rebound.

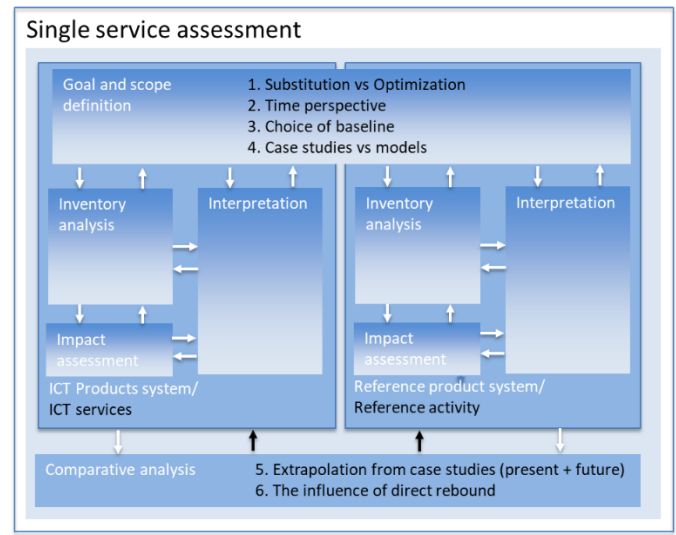

**Figure 1: Single service assessment framework according to the ETS/ITU standard modified in accordance with *Part I*. *Part I* enhancements are represented with numbered black text and symbols; unnumbered black text shows differences in terminology between this paper and the ETSI/ITU framework.**

The ETSI/ITU standard only considers the assessment of a single service. On a *multi-service* level – the focus of this paper – there is no standard for measuring the effects of ICT. The ITU L.1420 standard [22], which is aligned with the ISO [23] and the GHG protocol [24] standards, focuses on the direct and value-chain environmental impacts of ICT companies. Although it briefly mentions the possibility for companies to, on a voluntary basis, list "organizational activities to reduce GHG emissions", it provides no guidance for this. Nevertheless, this standard will be relevant to our work in defining a company's boundaries, as shown in Section 4.1.

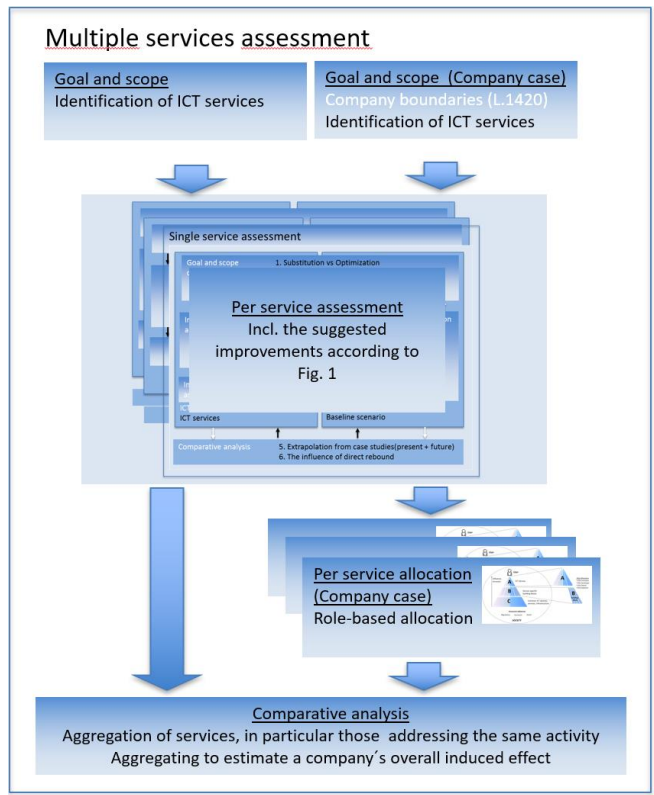

**Figure 2: This paper's suggested extension of the ETSI/ITU framework towards multiple services. The left part shows the assessment procedure for multiple services, the right part additionally includes the activities needed to assess multiple ICT services for a company.**

Fig. 2 outlines the contributions of our work in relation to the ETSI/ITU framework and the *Part I* [1] enhancements for single service assessments. On a multi-service level, the identification of services, the allocation of the environmental effects between them, and their aggregation are covered. The paper also discusses two aspects important for assessing the overall effect induced by a company's ICT services: the identification of company boundaries, and the allocation between several actors, contributing to each service.

## 4  THE MULTI-SERVICE AND COMPANY PERSPECTIVES

The single-service assessment (described in *Part I* [1]) had its starting point in the ETSI/ITU standard. For the multi-service case, there is no established correspondent. However, on top of the complex assessment of individual ICT services outlined in *Part I*, several challenges are specific to estimating the effects induced by a bundle of services. Some of these challenges are relevant to any multi-service assessment, for instance when assessing the potential effect of ICT services for smart sustainable cities [25], or for society-wide assessments such as those attempted by the GeSI studies [6-8]. For the special case of a company that wants to estimate the effects induced by its own ICT services, additional considerations apply.

This section starts by discussing the identification of relevant ICT services (Section 4.1). It further describes two possible sources for double counting: between several services addressing the same emissions (Section 4.2), and – specifically for companies devising the effects induced by their ICT services – between several companies contributing to the same service, and thus claiming the same reductions (Section 4.3). Finally, 4.D addresses the aggregation across services at company level.

### 4.1  Identification of ICT services

To be accurate, the choice of relevant ICT services must be specific (only consider relevant services) and complete (i.e., consider all relevant services) within the declared assessment boundaries (e.g. a company). Regarding completeness, the mechanisms leading to the induced effect of an ICT service were discussed in Part I [1]. Within the assessment boundaries, ICT services must be analysed whether they substitute or optimize other societal activities and might thus lead to an induced environmental effect. For companies, specifically, it is necessary to consider the company boundaries to identify the ICT services that belong to the company. For this, we refer fully to the principles outlined in [22].

Enabling ICT services must not necessarily have been designed with an explicit environmental goal. The environmental impact may appear as a side-effect, as the Kenyan "M-PESA" mobile money service [26] shows. Developed as a business solution, it also reduces bank-related travel and thus emissions.

As for specificity, only services having ICT as a key enabler [27] should be considered as ICT services. In particular, we agree with [5] that embedded microprocessor systems (e.g., motor optimization systems) and systems where ICT is mainly used as a tool for administration, design, or control (e.g., building design, large-scale renewables) should not be considered as ICT services. As an example, induced effects from the introduction of large-scale wind turbines are not due to ICT – although a particular ICT service that increases their efficiency may be. Section 6 addresses alternatives that might allocate only part of the induced effect to ICT, for services where ICT is one of many technologies involved. The choice of services for future assessments is more challenging as new (unpredictable) services might appear.

### 4.2 Aggregation of services, in particular those addressing the same reference activity

When aggregating the induced effect of ICT services that 'compete' for reducing the same GHG emissions, there is a risk that effects are inadvertently double counted. This happens, for example, if several of the services considered target behavioural changes of the same group of users, thereby impacting the same emissions through different mechanisms. This type of double counting is one of the criticisms of the overall ICT enablement potential from the much cited "SMARTer 2020" study [7], as analysed in [5]. The ITU study on the enablement potential of ICT in Korea 2011 and 2020 [28] illustrates the risk. The study identifies 14 potentially enabling ICT services. Next to the real-time navigation (RTN), one of the other services is the GPS-based real-time bus information system (BIS) that optimizes bus traffic convenience. RTN's potential is to reduce the time (and fuel) spent in traffic jams among car commuters and thus fuel consumption, while the BIS is expected to convince more commuters to change to public transports. The two ICT services compete to reduce the emissions from car commuting by modifying the same reference activity i.e. the car commuting, illustrating how double counting could occur unless considerations are taken to make sure that the baseline of the second service considers the reduction in overall emissions that would already be made through the first service.

When two or more services interact by competing to reduce the emissions of the same reference activity, their individual effects cannot simply be added. Once one modification has been applied, the next one has only a smaller footprint left to modify, the third one an even smaller one, and so on. The aggregated effect of n services $S_1..S_n$ modifying the same reference activity $A_k$ must then be computed via the residual footprint of the original reference activity $A_k$ after applying each modification in turn, as shown in Eq. 18[1]. This sequential approach leads to a smaller, more accurate, overall effect than if simply adding the individual effects:

$$E_X(S_1 \cup ... \cup S_n | A_k) = FP(A_k) * \left(1 - \prod_{i=1}^{n}\left(1 - \frac{E_X(S_i|A_k)}{FP(A_k)}\right)\right) \quad (18)$$

If Eq. 18 includes services that do not modify reference activity $A_k$, $E_X(S_i|A_k)$ of these services becomes zero and the term inside the product will equal 1; the service will thus not impact the multiplication result. This observation will be relevant for Eq. 20 below. If only one service $S_i$ modifies reference activity $A_k$, Eq. 18 reduces as expected to $E_X(S_1 \cup ... \cup S_n | A_k) = E_X(S_i|A_k)$.

Thus, when more services modify a reference activity, their joint effect is typically smaller than the sum of the individual effects would be. Sometimes, estimating their overall effect as shown in Eq. 18 suffices. Often, however, it is also relevant to derive the contribution from each individual service to the overall induced effect. For this purpose, we introduce $E_X^*(S_i|A_k)$, the individual service's share of the joint effect, which we propose to be proportional to its individual contribution:

$$E_X^*(S_i|A_k) = \frac{E_X(S_i|A_k)}{\sum_{l=1}^{n} E_X(S_l|A_k)} * E_X(S_1 \cup ... \cup S_n | A_k)$$
$$= \frac{E_X(S_i|A_k)}{\sum_{l=1}^{n} E_X(S_l|A_k)} * FP(A_k) * \left(1 - \prod_{i=1}^{n}\left(1 - \frac{E_X(S_i|A_k)}{FP(A_k)}\right)\right) \quad (19)$$

As for Eq. 18, any services $S_1..S_n$ that do not modify $A_k$, will not impact the result. Moreover, when only one $S_i$ modifies a certain reference activity $A_k$, it receives as expected the entire induced effect, as Eq. (19) reduces to $E_X^*(S_i|A_k) = E_X(S_i|A_k)$.

From the definition of $E_X^*(S_i|A_k)$ follows that the summation of all $E_X^*(S_i|A_k)$ equals the total effect of services impacting reference activity $A_k$, $E_X(S_1 \cup ... \cup S_n | A_k)$:

$$E_X(S_1 \cup ... \cup S_n | A_k) = \sum_{i=1}^{n} E_X^*(S_i|A_k) \quad (20)$$

---
[1] The Equations are continuing the number series from Part I to enable them being brought together in Section V.

Until Eq. 17, the equations have assumed a bijection between modified activities and modifying services, while Eq. 18 introduces the case of several ICT services modifying the same reference activity. In the most general case, the induced effect emerging from a combination of different ICT services $S_i \in \{S_1..S_n\}$ modifying several activities $A_k \in \{A_1..A_m\}$ can be assessed according to Eq. 21. As discussed above, in the calculation all services can be considered for each reference activity; for the activities they do not modify, their effect $E_X(S_i|A_k)$ equals zero.

$$E_X(S_1..S_n | A_1..A_m) = \sum_{k=1}^{m}\left(FP(A_k) * \left(1 - \prod_{i=1}^{n}\left(1 - \frac{E_X(S_i,A_k)}{FP(A_k)}\right)\right)\right) \quad (21)$$

Of course, Eq. 21 is more of a theoretical construct. In practice, we expect Eq. 18 to be used for each reference activity $A_k \in \{A_1..A_m\}$ separately, followed by an allocation to the related modifying services according to Eq. 19. Once these steps have been taken, the joint effect can be computed directly according to Eq. 22, which expands Eq. 20 to multiple activities.

$$E_X(S_1..S_n | A_1..A_m) = \sum_{i=1}^{n}\sum_{k=1}^{m} E_X^*(S_i|A_k) \quad (22)$$

## 4.3 Allocation between companies contributing to an ICT service

In theory, a company's induced environmental effect is the sum of the individual effects of the ICT services it offers, computed according to either Eq. 21 or Eq. 22. Typically, however, reality is far more complex. Rarely does a company contribute alone to a service; several companies and further stakeholders are usually involved in its development, installation, maintenance, and uptake. As more companies start to make emission reduction claims, this complex web of actors needs to be studied to see *who* could make a claim and *to what extent*. For assessing a company's overall induced effects, we thus revert to the assessment of single services as reflected in Eq. 19, analyse how these can be distributed among the various actors contributing to each service, and only after that aggregate across the different services of the company.

Although the induced effect of services should be assessed on a factual basis, a choice of allocation principles is also a matter of values. Without further analysis, [18] lists the following possible attribution approaches: i) equal allocation between 'all different elements' (i.e., actors), ii) financial cost attribution, iii) financial value attribution, iv) stakeholder consensus. When discussing different options below, we disregard the small conceptual difference between ii) and iii), omit iv) as we believe that stakeholder consensus is not a viable option in most practical cases, and add another principle in which only the main actor claims the entire effect:

- *The winner takes it all*: The induced effect can be claimed only by the company developing the ICT service used by the end-user. The service developer is indeed closest to the application and its effects; this company exercises the greatest influence and is arguably the least exchangeable player and is here acknowledged with a larger portion of the enabling effects in comparison with other principles. Besides its simplicity, this principle has the advantage that double counting among actors cannot occur, so sector-wide or country-wide aggregation of the effects of all services are, at least theoretically, straightforward. On the other hand, all the other actors contributing to the service, at the same or different levels, are not acknowledged.

- *Touch it and it's yours* is a quite generous approach that allows all companies that contributed to an ICT service to claim its full effect, including its downstream effects. Because numerous companies already report their value-chain footprint [22], it can be argued that it is only fair for them to also claim the downstream positive effects [18]. However, while value-chain reporting is about conservatively expanding responsibilities, in the enablement case the conservative approach is to restrain claims. Moreover, this principle seems a bit of a Pandora's box: how far behind in the supply chain of a service should the claim expire? Can it go as far down as the extraction of raw materials? It seems hard to link influence, effort and the induced effect. Such a principle might simply lead to credibility losses. Finally, it leads to double counting and does not allow for sector-wide aggregation. This paradigm is perhaps the most commonly adopted one due to the lack of well-founded alternatives.
- *Show me your money* tries to emulate the economic allocation frequently used in life cycle assessments. Here, credit should also go to all companies along the value chain proportionally to each company's costs or value added to the final product. Under this principle, double-counting is avoided, and aggregations seem possible, but complex. While appealing because it gives credit to all contributors, this principle also faces the question regarding how far back in the value chain it should reach. A further pragmatic issue is that product costs and benefits are often confidential. A third and more fundamental issue is that value added is not necessarily correlated to the importance of the contribution to the service. Someone's costless moment of inspiration, for example, might have been by far the most important contribution.

None of these principles satisfies the wish to include all involved actors while giving more credit to those with a more direct impact, and the lack of a consistent allocation approach is emphasized by companies such as AT&T [17]. The applicability of allocation principles applied by LCA studies was also evaluated as an option. However, such allocations were not helpful as our aim here is to allocate within a value-chain (between roles), whereas LCA allocations are primarily dealing with allocation of processes (between products) or between life cycles. We thus believe a new principle is needed.

The definition of a new allocation paradigm needs to deal with a complex reality where each ICT service relies on a large variety of equipment and supporting services. Unlike "touch it and it's yours", it must discern between the importance of the individual contributions, creating different levels of claims for the different qualities of contributions. We see three main contribution levels:

C. The main ICT service itself, which directly leads to an induced effect.
D. Dedicated building blocks (equipment or software), developed specifically for the A-level service.
E. General-purpose building blocks (equipment or software) required by the A-level service.

Starting from the least specific level, typical C-level ICT building blocks are telecom and computer networks (equipment and protocols), frontend devices such as smartphones or computers, and backend devices such as servers in data centers. The attribute best describing this level is necessary commonalities. Considering lower-level commonalities such as components or raw materials does not seem meaningful.

On B-level, the equipment or software must have been specifically built for the A-level ICT service. For a smart metering service, for example, the smart meter itself is such a B-level device. Likewise, a tablet application providing users with real-time information on their energy consumption is a B-level software component for smart metering. B-level equipment and sub-services typically use one or more C-level ICT building blocks. In addition to necessary, B-level components are specific.

Level A contains the overall ICT service that brings together all the B-level building blocks into one integrating service. Such a service typically makes use of several building blocks from both B- and C-level. Smart metering, for example, is a service that uses smart meter devices (B), backend data centers (C), a user feedback app (B), a network transmission protocol (C), a billing app (B), and several more such components. Level A is the integrating level.

It is clear from looking at the ICT service value chain that any proposed allocation principle must take these complexities into account. Considering the influence on the final usage, the A-level is closest to the induced effect, followed by B-level and a distant C-level; although all levels are necessary for the service. Furthermore, A- and B-level contributions can be more accurately allocated to particular actors. For these essential contributions, we consider that a "100% rule" must hold per level, which means that the sum of all enabling claims at that level should equal 100% of the total estimated induced effect. Double counting between actors is thus avoided. On C-level such principle does not seem practically feasible – for such generic building blocks it would be too challenging to identify all ICT services supported as well as the many individual actors contributing to those. For C-level, we propose the "touch it and it's yours" principle, while noting that an enablement statement on C-level is less specific and perhaps less useful than an enablement statement on A- or B-levels. Theoretically, an allocation could also be made between A- and B-levels to allow for aggregation of the two without double counting. This is further discussed in Section 4.4.

To apply the 100% rule, the allocation principle needs to consider both the different building blocks associated with each level – for B-level, its specific equipment and software blocks, for A only the ICT service itself – as well as the various stakeholders contributing to them. The first allocation step, only needed for the B-level, is to allocate between the specific building blocks needed for the service. As first approach, assuming all building blocks to be necessary for the service to function as intended, we propose that each block gets an equal share.

In a next step, the various stakeholders contributing to the ICT service at A-level, and to each building block at B-level, are considered. For each of them, these stakeholders can be:

I. The innovator (IN)
II. The developer (DE)
III. The service owner (OW)
IV. The operator (OP)

All of these actors are essential to the existence of the ICT service and thus to its environmental benefits: on the creative part, the innovating company, whose brainchild the service is, has arguably an outstanding role. So does the developer, who may or may not coincide with the innovator. In many cases, it might not be possible to pinpoint one innovator – e.g., for a standardized product relying on numerous patents. In such a case, the innovator role would be attributed to the developer. Finally, the company that bought the service and the one operating it are also essential for its existence.

The user is also an essential actor here – the one ultimately influencing the service usage. The user, however, whether an individual or a private or public organisation, is different from the stakeholders contributing to the service. Private users are not expected to make public environmental claims, while company users will take advantage of the reduction through the reduced footprint enabled by the service. The user is thus not part of the allocation of reductions induced by ICT, a view also held by [18].

For actors I-IV, an allocation that complies with the 100% rule must be established. Seeing all these actors as essential for the service, as a first approach we suggest sharing the benefits equally, as shown in Fig. 3. Depending on the service, some of these roles may coincide and the corresponding shares aggregated. A utility company (the owner), for example, could have bought a smart metering service from an ICT integrator (the developer). The service might be operated by the utility company (roles III and IV) or by the developer (roles II and IV). Obviously, roles are not fixed for a company, but different roles may apply for different services.

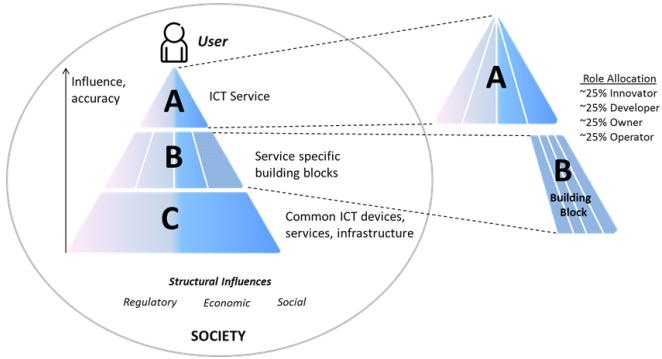

**Figure 3: The role-based allocation principle.**

Based on these observations, we define the following rules for allocating the induced environmental effect among actors I-IV:

1. The default allocation between roles I-IV is 25% each.
2. If no distinct innovator can be identified, that role is attributed to the developer.
3. If a stakeholder plays several roles for an ICT service, it can claim the cumulated percentages of its roles.

The argument for using equal shares between the stakeholders is not very elaborated, but more of pragmatic nature – all the roles are needed for the service to take off. The important message here is not to state an optimal allocation coefficient, but to identify the stakeholders, and to agree that the aggregated effect should be 100%. Section 6 shortly addresses econometric analysis as a possible alternative for determining the shares.

The effect induced by an ICT service $S_i$ modifying reference activity $A_k$, possibly jointly with other services, $E^*_X(S_i|A_k)$, is computed according to Eq. 19. The induced A-level effect per role $Ro$ is derived from splitting this effect equally between the four roles:

$$EA^*_{X,Ro}(S_i|A_k) = \frac{1}{4} * E^*_X(S_i|A_k) \text{ for } Ro \in \{IN,DE,OW,OP\} \quad (23)$$

As a next step, the induced effects of the A-level roles held by the assessed company (typically not all of them) are added together to calculate the total A-level contribution of the company for service $S_i$, $EA^*_{X,Ro}(S_i|A_k)$.

$$EA^*_{X,Co}(S_i|A_k) = \sum_{Ro} EA^*_{X,Ro}(S_i|A_k) \quad (24)$$

for all roles Ro ∈ {IN,DE,OW,OP} taken by company Co for the service $S_i$.

Similarly, the effect on B-level needs to be split among its building blocks, $BB_j$, and then among the four roles of each building block. For each B-level building block $BB_j$, the induced effect per role, $EB^*_{X,Ro,BB_j}(S_i|A_k)$, becomes:

$$EB^*_{X,Ro,BB_j}(S_i|A_k) = \frac{1}{|BB|} * \frac{1}{4} * E^*_X(S_i|A_k),$$
$$\text{for } Ro \in \{IN,DE,OW,OP\} \quad (25)$$

where |BB| represents the number of B-level building blocks.

To aggregate a company's total B-level effect for a service $S_i$, first the induced effects of the roles the company holds are added together for each building block, and then aggregated across all building blocks (if the company contributed to more than one):

$$EB^*_{X,Ro}(S_i|A_k) = \sum_{BB_j} \sum_R EB^*_{X,Ro,BB_j}(S_i|A_k) \quad (26)$$

for the company-relevant roles Ro ∈ {IN,DE,OW,OP}.

Finally, according to the "touch it and it's yours approach", each C-level contributor may make the less specific claim of contributing to the entire induced effect:

$$EC^*_{X,Ro}(S_i|A_k) = E^*_X(S_i|A_k) \quad (27)$$

Eqs. 23-27 start from the induced effect of an individual service as represented by the service's share of the joint effect according to Eq. 19, which reflects a generalized situation in which several services might modify the same reference activity. If each reference activity is modified by one service only, $E^*_X(S_i|A_k) = E_X(S_i|A_k)$, as discussed in Section 4.2, and $E_X(S_i|A_k)$ can replace $E^*_X(S_i|A_k)$ in Eqs. 23, 25, and 27.

### 4.4 Estimating a company's overall induced effect

As Eqs. 23, 25 and 27 are based on $E_X^*(S_i|A_k)$ and thus avoid double counting, the final aggregation of the assessed ICT services is a straightforward addition, whether these services modify the same reference activity or not. The A-, B- and C-level effects induced by company Co in point X, $EA_{X,Co}(S_1..S_n|A_1..A_m)$, $EB_{X,Co}(S_1..S_n|A_1..A_m)$, and $EC_{X,Co}(S_1..S_n|A_1..A_m)$, respectively, are derived by simply adding the company's contributions across all its services and the activities they modify, in line with Eq. 22:

$$EA_{X,Co}(S_1..S_n|A_1..A_m) = \sum_{i=1}^{n}\sum_{k=1}^{m} EA_{X,Ro}^*(S_i|A_k) \quad (28)$$

$$EB_{X,Co}(S_1..S_n|A_1..A_m) = \sum_{i=1}^{n}\sum_{k=1}^{m} EB_{X,Ro}^*(S_i|A_k) \quad (29)$$

$$EC_{X,Co}(S_1..S_n|A_1..A_m) = \sum_{i=1}^{n}\sum_{k=1}^{m} EC_{X,Ro}^*(S_i|A_k) \quad (30)$$

It is important to keep in mind that A-, B- and C-levels are addressing different layers of the same systems and thereby of the same effect and cannot be added together without double counting.

A further step along our paradigm would be to avoid double counting entirely, by allocating the effect between the different levels. Such paradigm would not be a difficult conceptual step, but would need an allocation key between the levels A and B, and possibly the elimination of any C-level claims entirely.

The authors are unsure about the latter, and even more so whether an allocation between the A- and the B-level exists that can be reasonable for the variety of existing services. In the context of increasing and undifferentiated claims (many of them on C-level), the focus of our paper was rather to conceptualize the existence of different contribution levels and to distinguish between them in a reasonable way. We leave further refinement to the community. At his stage we only note that current claims tend to mix contributions at different levels, and follow *a touch it and it´s yours* paradigm. Without the establishment of common and widely adopted allocation practices based on principles such as the ones described in this section, we do not agree with [18] that the aggregation to an investment portfolio is a relatively simple, or even feasible, task.

Furthermore, the sector-level aggregation is outside the scope of our article but its relevance depends on whether the allocation principle respects the 100% rule or not. For the role-based allocation, A- and B-levels can each be aggregated to a sector level but not added together unless an allocation is made between them; and in any case C-level cannot.

## 5 SUGGESTED WORKFLOW FOR DERIVING THE INDUCED EFFECT OF A COMPANY´S ICT SERVICES

As making company-level enablement claims is a common practice, we propose a methodological workflow which supports companies to apply the methodology proposed in *Part I* (Bullets 1, 3) and in this paper (Bullets 2, 4, 5) in assessing the induced effect of their services.

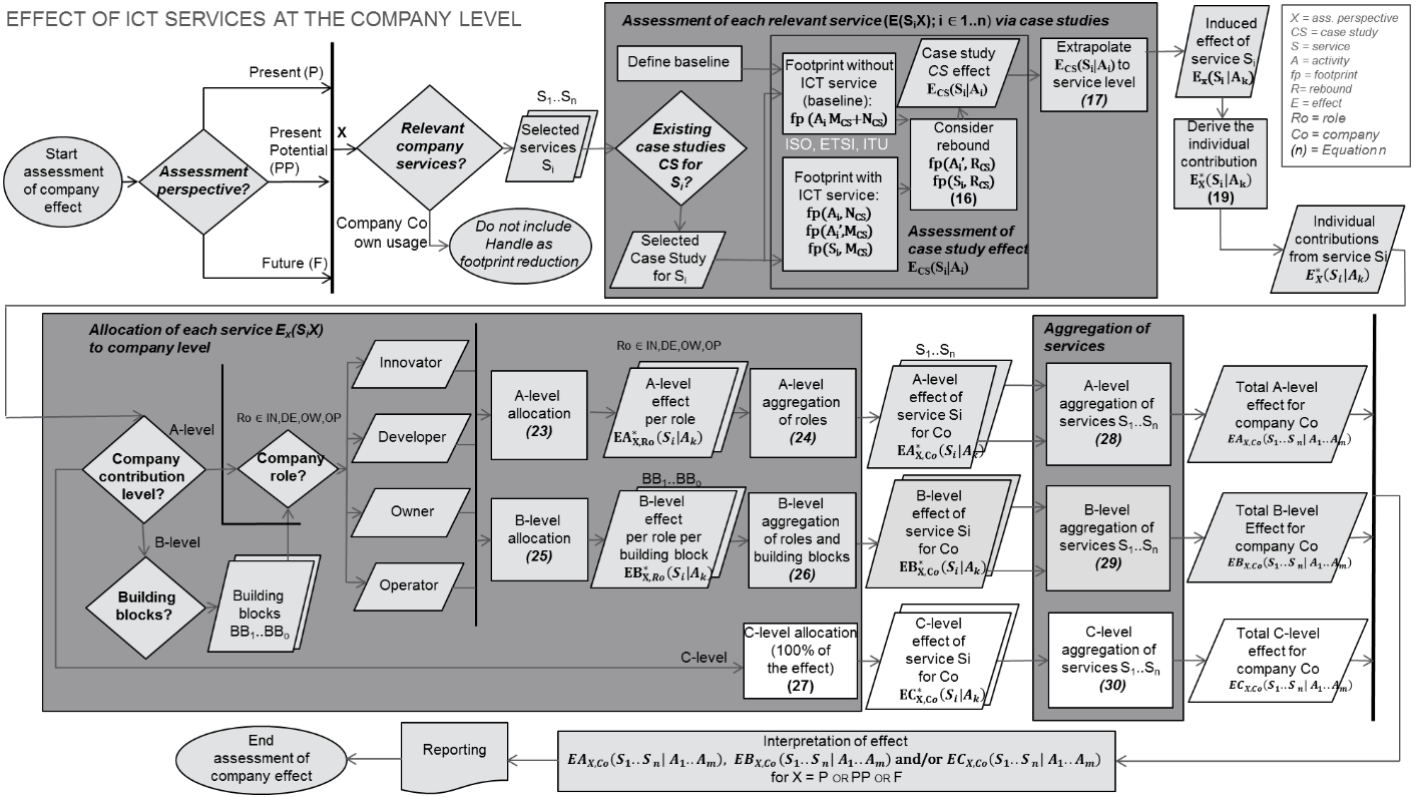

**Figure 4: Workflow for assessing the induced effect of ICT companies.**

A. The first step defines the time perspective – does the study assess the present situation (P), the circumstantial potential of the present (PP), or a future potential (F)?
B. The second step describes the choice of relevant ICT services: which services can be claimed by an ICT company?
C. The third methodological step assesses the induced effect of each relevant service. After defining the baseline for the reference activity and taking direct rebound effects into consideration, the induced effect is computed by conservatively extrapolating from case studies (Eq. 17), and its individual effect among other ICT services is derived (Eq. 19).
D. The fourth step allocates the induced effect of each service to the actors contributing to it. Per service, this means identifying the level(s) of contribution of the company (A, B, and/or C), and allocating, on each level, role-specific contributions according to Eqs. 23-27.
E. Finally, the last step aggregates all induced effects of one company. The aggregation is done per A-, B-, and C-level in a straightforward way (Eqs. 28-30).

The overall workflow, based on the use of case studies, is outlined in Fig. 4.

## 6. DISCUSSION AND FUTURE RESEARCH

The discussion in *Part I* addressed whether there is a need for enhanced assessment methods for the induced effects of ICT, as well as considerations around the hypothetical baseline and uncertain extrapolations. Here, we discuss allocation principles (Section 6.1) and the extrapolation of our method to other types of impacts (Section 6.2). For these aspects, we identified the issue but, given the breadth of the analysis and the complexity of the topics, we only proposed a first solution, which is often simple and pragmatic. More generally, as stressed in *Part I*, the proposed methodology should also be more thoroughly tested to confirm its usability for complex services and scenarios.

### 6.1 Allocation principles

Section 4.1 argued to only consider services that include ICT at the very core of their solution and ignore systems using ICT merely as a tool for administration, design or control. This basic and conservative principle protects against gross overestimates due to exaggerated allocation to the ICT sector. It could be improved, however, by a more refined allocation principle between the ICT sector and other sectors. Likewise, the intra-sectorial allocation of 25% to each essential actor suggested in Section 4.3 could also be further developed. Devising a more refined allocation principle for these two instances is beyond the scope of this paper. A good starting point, however, might be Solow's growth accounting [29], which addresses a similar problem by decomposing economic growth down to its different influencing factors and has been econometrically modelled as a linear regression model [30].

The usage of ICT services is evolving over time. This dynamicity is implicitly reflected in the equations through the introduction of present, present potential and future. The allocation between the different roles (particularly OW and OP) may also change over time, as may the average per-usage effect. Strictly spoken, these dynamic perspectives should be included more explicitly in the equations, even when considering the yearly emissions and not the full life time. However, as the equations are intended to outline the principles, rather than giving exact calculation instructions, such modifications would bring too much complexity at this stage.

A fundamental difficulty in assessing the environmental effect of ICT stems from the uncertain development of policy, other technology, and user behavior, e.g. related to the social embedment and the exceptional dynamics of innovation and diffusion as mentioned in [27]. These will impact the baseline for the reference activity, the future use of ICT, and rebound. Specifically, for baselines, the potential impact from policy and other technology is perhaps most evident for a projection-based baseline (case iii) in Fig. 3a of *Part I*. In principle, a strong development of technology could lead to point iii) being placed far below point i) – a baseline fixed according to the conditions at the introduction of the ICT service. This would have been the case e.g. if looking at NOx-emission projections at the time of the introduction of catalytic converters in cars. Similarly, the baseline can also be affected by policy changes - a baseline can look quite differently with or without accounting for the introduction of various environmental taxes or regulations. The importance of policy for a study of the future impact of an ICT service is thus crucial. Policy can directly support or counteract a specific service and that policy can entirely change the context in which a service is expected to function. As challenging as it appears to consider such factors, the first step is to transparently list any assumptions made in this direction.

### 6.2 Extrapolating the principles to other impact types

The article focused on GHG emissions. This measure was taken for simplicity, but many of the topics addressed seem equally relevant to other environmental impact categories. Applying the assessment principles postulated here to any other type of environmental impact would be straightforward. Furthermore, the entire frame of thoughts could provide input for assessing the socially enabling effects of ICT services, and could show how companies contribute to the Sustainable Development Goals [31] for which ICT is expected to play a major role [32]. Further research is needed to investigate such an extended use of the principles defined here.

## 7 CONCLUSION

Starting from the ETSI/ITU framework with suggested enhancements in line with *Part I*, this article identified challenges for understanding the environmental effects induced by multiple ICT services through substitution or optimization of reference activities. Beyond identifying common flaws in existing assessments, the article put forward solutions to help establish a more rigorous and comprehensive methodology for assessing the induced effects of several ICT services, in particular by companies wanting to understand the effects induced by their services.

On this multi-service level, the focus was on the identification of ICT services and aggregation of services addressing the same emissions. For the special case of company claims, the main challenge addressed was the allocation between all actors contributing to a service. A novel allocation solution was proposed which con-siders layers of contributors with different roles. Solutions are also proposed for the subsequent aggregation of services in order to estimate a company´s overall induced effect. Our contribution offers guidance to practitioners, making them aware of the common pitfalls, and of principles to avoid them. This guidance is summarized into a workflow for company-level assessments. Although our methodology does not provide a cookbook recipe for all steps along the way, the conceptualization should increase awareness regarding the complexities of assessing induced effects of multiple ICT services.

## ACKNOWLEDGEMENTS

The authors wish to thank Dr. Åsa Moberg for general feedback on earlier versions of this article. The authors are also thankful for the support from Prof. Radu Craiu regarding statistical methods for volunteer biases, and Dr. Dragoș Radu regarding econometric models for discriminating among influencing factors. This research was financed by the organizations of the authors and with co-funding for the KTH-part from Vinnova, Sweden's innovation agency.

# APPENDIX: EQUATION OVERVIEW

This appendix collects the different equations from Part I [1] and Part II (this paper) of the article.

**Table 1. Equation overview**

| | | |
|---|---|---|
| (1) | $E(S_i|A_i) = FP(A_i) - FP(S_i)$ | Basic ETSI/ITU equation |
| (2) | $E(S_i|A_i) = (FP(A_i, M)+FP(A_i, N)) - (FP(A_i, N) + FP(S_i, M)) = FP(A_i, M) - FP(S_i, M)$ | Introduction of partial substitution |
| (3) | $E(S_i|A_i) = (FP(A_i, M)+FP(A_i, N)) - (FP(A_i, N) + FP(A_i', M) + FP(S_i, M)) = FP(A_i, M) - (FP(A_i', M) + FP(S_i, M))$ | Introduction of optimization |
| (4) | $E_X(S_i|A_i) = (FP(A_i, M_X)+FP(A_i, N_X)) - (FP(A_i, N_X)+FP(A_i', M_X)+FP(S_i, M_X)) = FP(A_i, M_X) - (FP(A_i', M_X) + FP(S_i, M_X))$ | Introduction of time perspective |
| (5) | $E_X(S_i|A_i) = \tilde{e}_{mod}(S_i|A_i) * |M_X|$ | Introduction of modelling approach |
| (6) | $E_X(S_i|A_i) = \tilde{e}_{CS}(S_i|A_i) * |M_X|$ | Introduction of case study approach |
| (7) | $e_j(S_i|A_i) = fp_j(A_i) - ((fp_j(A_i')+fp_j(S_i))$ | Introduction of per-usage effect |
| (8) | $E_{CS}(S_i|A_i) = \sum_{j \in M_{CS}} e_j(S_i|A_i)$ $= \sum_{j \in M_{CS}}(fp_j(A_i) - (fp_j(A_i') + fp_j(S_i)))$ | Summing up per-usage effects to a case study effect |
| (9) | $\tilde{e}_{CS}(S_i|A_i) = E_{CS}(S_i|A_i) / |M_{CS}| = (\sum_{j \in M_{CS}}(fp_j(A_i) - (fp_j(A_i')+fp_j(S_i)))) / |M_{CS}|$ | Calculating average per-usage case study effect |
| (10) | $E_X(S_i|A_i) = \tilde{e}_{CS}(S_i|A_i) * |M_X| = E_{CS}(S_i|A_i) * |M_x|/|M_{CS}| = (\sum_{j \in M_{CS}}(fp_j(A_i) - (fp_j(A_i')+fp_j(S_i)))) * |M_x|/|M_{CS}|$ | Extrapolation of a case study to a service level |
| (11) | $E_X(S_i|A_i) = k_X * \tilde{e}_{CS}(S_i|A_i) * |M_X|$ | Expanding (6), the introduction of the case study approach (also reflected in the first part of (10)), with case study quality coefficient |
| (12) | $E_X(S_i|A_i) = k_X * E_{CS}(S_i|A_i) * |M_x| / |M_{CS}|$ $= k_X*(\sum_{j \in M_{CS}}(fp_j(A_i) - (fp_j(A_i') + fp_j(S_i)))) *|M_x|/|M_{CS}|$ | Expanding second and third part of (10) with case study coefficient |
| (13) | $E_X(S_i|A_i) =$ $FP(A_i, M_X) - FP(S_i, M_X + R_X)$ | Introducing rebound for substitution |
| (14) | $E_X(S_i|A_i) =$ $FP(A_i, M_X) - (FP(A_i', M_X + R_X) + FP(S_i, M_X + R_X))$ | Introducing rebound for optimization |
| (15) | $E_X(S_i|A_i) =$ $FP(A_i, M_X + R_X) - (FP(A_i', M_X + R_X) + FP(S_i, M_X + R_X))$ | Illustration of the (non-correct) overstated effect |
| (16) | $E_{CS}(S_i|A_i) = \sum_{j \in M_{CS}}(fp_j(A_i) - (fp_j(A_i') + fp_j(S_i))) - \sum_{j \in R_{CS}}(fp_j(S_i) + fp_j(A_i')) = \sum_{j \in M_{CS}}(fp_j(A_i)) - \sum_{j \in (M_{CS}, R_{CS})}(fp_j(S_i) + fp_j(A_i'))$ | Expanding case study effect (8) with direct rebound |
| (17) | $E_X(S_i|A_i) =$ $k_X * (\sum_{j \in M_{CS}}(fp_j(A_i)) - \sum_{j \in (M_{CS}, R_{CS})}(fp_j(S_i) + fp_j(A_i'))) *$ $|M_X| / |M_{CS}|$ | Expanding calculation of the induced effect at a service level (12) derived from a case study with direct rebound |
| (18) | $E_X(S_1 \cup ... \cup S_n|A_k) =$ $FP(A_k) * \left(1 - \prod_{i=1}^{n}\left(1 - \frac{E_X(S_i|A_k)}{FP(A_k)}\right)\right)$ | Gradual aggregation for several services modifying the same activity |
| (19) | $E_X^*(S_i|A_k) = \frac{E_X(S_i|A_k)}{\sum_{l=1}^{n} E_X(S_l|A_k)} *$ $E_X(S_1 \cup ... \cup S_n|A_k)$ $= \frac{E_X(S_i|A_k)}{\sum_{l=1}^{n} E_X(S_l|A_k)} * FP(A_k) *$ $\left(1 - \prod_{i=1}^{n}\left(1 - \frac{E_X(S_i|A_k)}{FP(A_k)}\right)\right)$ | Deriving the individual service's share of the joint effect on one activity |
| (20) | $E_X(S_1 \cup ... \cup S_n|A_k) = \sum_{i=1}^{n} E_X^*(S_i|A_k)$ | Deriving the overall effect on one activity through adding individual contributions of each service to get the joint effect |
| (21) | $E_X(S_1..S_n | A_1..A_m) = \sum_{k=1}^{m}(FP(A_k) *$ $\left(1 - \prod_{i=1}^{n}\left(1 - \frac{E_X(S_i, A_k)}{FP(A_k)}\right)\right))$ | Gradual aggregation for several services modifying a set of activities |
| (22) | $E_X(S_1..S_n | A_1..A_m) =$ $\sum_{i=1}^{n} \sum_{k=1}^{m} E_X^*(S_i|A_k)$ | Deriving the overall effect on a set of activities through adding the individual contributions of each service to calculate the joint effect |
| (23) | $EA_{X,Ro}^*(S_i|A_k) = \frac{1}{4} * E_X^*(S_i|A_k)$ | Per-service allocation to A-level |
| (24) | $EA_{X,Co}^*(S_i|A_k) = \sum_{Ro} EA_{X,Ro}^*(S_i|A_k)$ | A-level role aggregation to service level |
| (25) | $EB_{X,Ro,BB_j}^*(S_i|A_k) = \frac{1}{|BB|} * \frac{1}{4} * E_X^*(S_i|A_k)$ | Per-service allocation to B-level |
| (26) | $EB_{X,Ro}^*(S_i|A_k) =$ $\sum_{BB_j} \sum_{R} EB_{X,Ro,BB_j}^*(S_i|A_k)$ | B-level role aggregation to service level |
| (27) | $EC_{X,Ro}^*(S_i|A_k) = E_X^*(S_i|A_k)$ | C-level role aggregation to service level |
| (28) | $EA_{X,Co}(S_1..S_n | A_1..A_m) =$ $\sum_{i=1}^{n} \sum_{k=1}^{m} EA_{X,Ro}^*(S_i|A_k)$ | Aggregation of A-level contributions to a company level |
| (29) | $EB_{X,Co}(S_1..S_n | A_1..A_m) =$ $\sum_{i=1}^{n} \sum_{k=1}^{m} EB_{X,Ro}^*(S_i|A_k)$ | Aggregation of B-level contributions to a company level |
| (30) | $EC_{X,Co}(S_1..S_n | A_1..A_m) =$ $\sum_{i=1}^{n} \sum_{k=1}^{m} EC_{X,Ro}^*(S_i|A_k)$ | Aggregation of C-level contributions to a company level |